
\input harvmac
\def\b{\bar}
\def\t{\tilde}
\def\h{N_h}
\def\g{N_g}
\font\cmss=cmss10 \font\cmsss=cmss10 at 7pt
\def\IZ{\relax\ifmmode\mathchoice
{\hbox{\cmss Z\kern-.4em Z}}{\hbox{\cmss Z\kern-.4em Z}}
{\lower.9pt\hbox{\cmsss Z\kern-.4em Z}}
{\lower1.2pt\hbox{\cmsss Z\kern-.4em Z}}\else{\cmss Z\kern-.4em Z}\fi}

\vsize=8.75truein
\hsize=5.5truein
\voffset=0.25truein
\hoffset=0.5truein
\nopagenumbers

\rightline{\vbox{\baselineskip12pt\hbox{FTUAM-93-07}\hbox{NEIP93-001}}}
\vskip2pt
\bigskip\medskip
{\vbox{ \centerline{\bf BARYON ASYMMETRY, SUPERSYMMETRY AND}
\vskip2pt   \centerline{\bf GRAVITATIONAL ANOMALIES{$^*$}}}}
  \footnote{} {$^*$
  Talk presented by FQ at the Texas/Pascos Conference,
 Berkeley Dec. 1992.  }
\bigskip\bigskip

\centerline{\bf Luis E. Ib\'a\~nez{$^1$}
 and Fernando Quevedo{$^2$}}
 \medskip
  \centerline{\it {$^1$}Departamento de  F\'{\i}sica
Te\'orica C-XI}
\centerline{\it Universidad Aut\'onoma de Madrid, Spain}
\medskip
\centerline{\it {$^2$}Institut de Physique}\centerline{\it Universit\'e de
 Neuch\^atel, CH-2000, Switzerland}

 \vskip .2in\bigskip\bigskip
 \noindent
\vbox{\baselineskip12pt
\centerline{\bf Introduction}
\medskip\medskip
We discuss two independent issues about the baryon asymmetry of
the universe. First, assuming that it is generated by
an unspecified source at high temperatures, we study the effects
of non-perturbative $SU(2)_W$ dynamics above the electroweak scale,
in the context of supersymmetric models. We find  that there is a
substantial difference with the nonsupersymmetric case with the net
effect of relaxing previous bounds on $B$ and $L$ violating
interactions. In particular supersymmetry allows neutrino masses as
large as  $10$ eV, consistent with solar neutrino data as well as what
is required for hot dark matter (together with cosmic strings or cold
dark matter) to explain COBE results on large scale density
perturbations. It is also consistent with neutrino oscillations
measurable at future accelerator experiments.
Second, we argue that the existence of a mixed lepton
number--gravitational anomaly in the standard model will induce $B-L$
violating interactions. These transitions would be catalized by
Einstein--Yang--Mills instantons or sphalerons and could create a
primordial $B-L$ asymmetry at Planck temperatures or lower. Gravity
(and the anomaly structure of the standard model) could then be the
ultimate source of the baryon asymmetry. We analyze the viability
of the presently known gravitational instantons and
sphalerons to realize this scenario. }


\lref\man{N. S. Manton, Phys. Rev. D28 (1983) 2019\semi
F. R. Klinkhammer and N. S. Manton, Phys. Rev. D30 (1984) 2212.}
\lref\krs{V. A. Kuzmin, V. A. Rubakov and M. E. Shaposhnikov, Phys. Lett.
155B (1985) 36\semi
P. Arnold and L. McLerran, Phys. Rev. D36 (1987) 581; D37 (1988) 1020.}
\lref\ya{M. Fukugita and T. Yanagida, Phys. Rev. D42 (1990) 1285.}
\lref\hatu{J. A. Harvey and M. S. Turner, Phys. Rev. D42 (1990) 3344}
\lref\cdeo{B. A. Campbell, S. Davidson, J. Ellis and K. A. Olive,
Phys. Lett. 256B (1991) 457; CERN-TH.6208/91 (1992).}
\lref\fglp{W. Fischler, G.F. Giudice, R.G. Leigh and S. Paban, Phys.Lett.
258B (1991) 45.}
\lref\barr{S.M. Barr, R.S. Chivukula and E. Fahri, Phys.Lett. B241 (1991)
 387
    \semi  S. M. Barr, Phys. Rev. D44 (1991) 3062\semi
D. Kaplan, Phys. Rev. Lett.  68 (1992) 741.}
\lref\ird{L. E. Ib\'a\~nez and G. G. Ross, Nucl. Phys. B368 (1992) 3.}
\lref\thoo{G. 't Hooft, Phys. Rev. Lett. 37 (1976) 8;
 Phys. Rev. D14 (1976) 3422.}
\lref\shapdin{ For a review and references, see M.E. Shaposhnikov,
CERN-TH.6304/91 (1991) \semi P.B. Arnold, Argonne preprint ANEL-HEP-CP-90
-95 (1990).}
 \lref\nbb{A.E. Nelson and S.M. Barr, Phys.Lett. B246 (1991) 141.}
\lref\kss{S.Y. Khlebnikov and M.E. Shaposhnikov, Nucl.Phys. B308 (1988)
885.}
\lref\msw{For a review see S. Mikheyev and A. Smirnov,
Sov.Phys.Usp. 30 (1987) 759.}
\lref\mswr{A. Dar and S. Nussinov, Particle World  2 (1991) 117 \semi
L.E. Ib\'a\~nez, `Beyond the Standard Model (yet again)', CERN-TH.5982/91
, to appear in the Proceedings of the 1990 CERN School of Physics \semi
R.N. Mohapatra, preprint UMD-PP-91-188 (1991) \semi
S.A. Bludman, D.C. Kennedy and P.G. Langacker, preprint UPR 0443T (1991)
\semi  M. Fukugita and T. Yanagida, Mod.Phys.Lett. A6 (1991) 645.}
\lref\zwir{G. Costa and F. Zwirner, Rivista del Nuovo Cimento
9 (1986) 1.}
\lref\mohap{R.N. Mohapatra, Nucl. Instrum. Methods A 284 (1989) 1.}
\lref\tak{M. Takita et al., Phys.Rev. D34 (1986) 902.}
\lref\iq{L.E. Ib\'a\~nez and F. Quevedo, Phys. Lett. B283 (1992) 261
and references cited therein. }
\lref\thd{G. 't Hooft, Nucl. Phys. B315 (1989) 517 .}
\lref\gibbons{G. Gibbons Ann. Phys. 125 (1980) 98,
 Phys. Lett. B84 (1979) 431. }
\lref\eh{T. Eguchi and A. Hanson, Ann. Phys. (N.Y.), 120 (1979) 82. }
\lref\bm{ R. Bartnik and J. McKinnon, Phys. Rev. Lett. 61 (1988) 141\semi
D. Galtsov and M. Volkov, Phys. Lett B273 (1991) 255. }
\lref\mz{ R. Mohapatra and X. Zhang, Phys. Rev. D45 (1992) 2699. }
\baselineskip12pt

 \vskip.2in\bigskip

 \centerline{\bf Global Symmetries in the Supersymmetric Standard Model}

 \medskip\bigskip

 We will consider the minimal supersymmetric extension of the
  Standard Model
 with gauge symmetry $SU(3)\times SU(2)\times U(1)$,
 three generations of quark ($Q, u_L^c, d_L^c$ ) and
 lepton ($L, E_L^c$)
 superfields, and two
 Higgs doublets  superfields ($H, \b H$). The gauge-invariant
 dimension-four operators that can appear in the superpotential are:
 \eqn\sup
{\eqalign{W=h_u\, Q_Lu_L^c\b H + h_d\, Q_Ld_L^c H+h_l\, L_L E_L^c H\cr
 + h_B\, u_L^c d_L^c d_L^c +h_L\, Q_Ld_L^cL+h'_L\, L_LL_LE_L^c \ ,\cr}}
where generation and gauge indices have been suppressed.
We can investigate the global symmetries of
the supersymmetric standard model (SSM)
        we just described. It is easy to see that there are
       only two symmetries   \iq.
A standard  one $g_{PQ}$ and a $R$ symmetry $I$.
 for which $(Q_L,u^c_L,d^c_L,L_L,E^c_L,\t H,\t{\b H})$
have the charges
$(0,-2,1,-1,2,-1,2)$  under $g_{PQ}$ and $(-1,-3,1,-1,1,-1,3)$,
under $I$.
Gauginos have zero charge
under $g_{PQ}$ and $1$ under $I$.
For $I$, the bosonic component of the  superfield
has one unit more than the corresponding fermion.
Both symmetries have mixed anomalies
for any number of generations $\g$ and Higgs  pairs   $\h$.
 It is convenient to define two independent
combinations $R_2\equiv I^{\g}\times g_{PQ}^{ 6-4\g}$ and $R_3\equiv
 I^{\g-\h}\times
g_{PQ}^{4+2\h-4\g}$ with the property that $R_N$ has only $SU(N)$
and $U(1)_Y$ mixed anomalies. The combination $B+L+R_2$ has no $SU(2)$
 nor $SU(3)$
mixed anomalies for the physical case $N_g=3$. Thus, just like
$B-L$, the symmetry $B+L+R_2$ is conserved

Let us now restrict to the case where all $B$-  or $L$-violating
terms in \sup\ are forbidden. Therefore there will be a total
of six global symmetries $B, L_i, R_2$ and $R_3$, where $R_2$
and $B+L$ have mixed $SU(2)$ anomalies and $R_3$ has
mixed $SU(3)$ anomalies. In order to explore the physical
 consequences of these anomalies we have to find the effective
operators that they generate. For the Standard Model the
effective operator from the $(B+L)- SU(2)^2$ anomaly is \thoo\
%
${O_1= (Q_L Q_L Q_L L_L)^{\g},}$
where the power of $\g$ is actually a product over
generations.
The analogous operator in the SSM is:
\eqn\opd{O_2=(Q_LQ_LQ_LL_L)^{\g}(\t H\t{\b H})^{\h}\t W^4\ \ ,}
where now the $\h$ higgsinos ($\tilde H, \tilde{\bar H}$) and
the winos ($\tilde W$) transforming non-trivially under
$SU(2)$  also contribute to the anomaly and then to \opd ,
in a way proportional to the Casimir of the corresponding
representation. For the
$R_3-SU(3)^2$ anomaly \mz, \iq\ the corresponding operator is
\eqn\opt{O_3=(Q_L Q_L u_L^c d_L^c)^{\g} \t g^6\ .}
%


The question we will address now is, given a primordial excess
of $B$ and $L$, how    they evolve if the reactions induced by
\opd\ and \opt\ are in thermal equilibrium,  i.e.  they occur
faster than the expansion rate of the universe parametrized by
the Hubble constant.                    For this we need to
express the $B$ and $L$ number density in terms of chemical
potentials. Using the constraint that
all the SSM interactions be in thermal equilibrium, we
will find only a few  independent chemical potentials
on  which depend all the number densities.      Extra constraints
are obtained if the anomalous processes \opd\ and \opt\ are
also in equilibrium.

For ultrarelativistic particles, the equilibrium number density
$\Delta n$
 (difference of particles and antiparticles) of a
particle species, depends on the
temperature $T$ and chemical potential $\mu$ of the respective
particles in the
following way:
\eqn\nude{{\Delta n\over s}=
{15\, g\over 4\pi^2 g_{*}}
              \big( {\mu\over T}\big )\cases{2\qquad{\rm bosons}\cr
              \noalign{\vskip2pt}
              1\qquad{\rm fermions}\,\cr}}
where $s={2\pi^2 g_{*}T^3 /    45}$ is the entropy density, $g$
is the
number of internal degrees of freedom and
$g_{*}$ is the total number of degrees of freedom ($\sim 200$
in the
supersymmetric case).
We will name the chemical potential of
a given particle by the name of the corresponding particle.
Since we will work at scales much higher than $M_W$,
all the particles in the same $SU(3)\times SU(2)\times U(1)$
multiplet have the same chemical potential and  the corresponding
gauge fields have vanishing chemical potential. Because of
generation-mixing interactions, the quark  potentials will be taken
generation-independent. Also, all the Higgses will
 have the same
chemical potential. Then, we have  to consider
$25=2\times 11 +3$ independent chemical potentials for the SM
particles (with two Higgses) plus their superpartners and the three
gauginos $\t W, \tilde g, \t B^0$.
The interactions in \sup\ imply
\eqn\thequ{u_L^c+Q_L+\b H  = 0\qquad
                   d_L^c + Q_L + H= 0 \qquad
                   E_L^{ci}+ L_L^i +H = 0\, .}
Gaugino couplings of the SSM imply also:
%
${\t Q_L = Q_L-\t g = Q_L - \t B^0 , }$
 with similar relations for the leptons, higgsinos and
                   right-handed quarks and leptons.
If all these processes are simultaneously in equilibrium, we
will have $\t B^0=\t W=\t g$.
We
are then left with the independent chemical potentials
$Q_L, L_L^i, H,\b H, \t g$.
We now  compute the total electric charge density $Q$ and
impose the constraint \hatu\ that it vanishes in a universe in  thermal
equilibrium. This implies
\eqn\charge{Q={
15\over 4\pi^2 g_{*} T}\{6\g(Q_L-L_L)+3(\h+2\g)(\b H - H)\}=0. }
Now we will use the condition that the $SU(2)$ and $SU(3)$
anomalous couplings
obtained from \opd\  and \opt\ are also in equilibrium at
high temperatures,
thus implying the relations
\eqn\aneqp{\eqalign{3\g Q_L+\g L_L+\h (H+\b H)+(4+2\h)\t g=0.\cr
  2Q_L+u_L^c+d_L^c+2\t g=0 \, .}}
The condition of vanishing electric charge, together with \aneqp,
 reduce the
number of independent variables to $Q_L$ and $\t g$.
We then find the expression
for the baryon and lepton densities,
\eqn\barlep{\eqalign{B&={30\over 4\pi^2 g_{*} T}
\{6\g\,Q_L-(4\g-9)\,\t g\} ,\cr
\noalign{\vskip2pt}
L&=-{45\over 4\pi^2 g_{*} T}\{ {{\g(14\g+9\h)}\over{\h+2\g}}{Q_L}
+\Omega(\g,\h) \t g\} }\  ,}
where $\Omega$ is an unimportant rational function.
Here  we have defined $L\equiv {1\over{\g}}\sum_{i} L^i$.
We can easily see that even setting $B-L=0$   we will get a
non-vanishing
$(B+L)\propto \t g\ne 0$, indicating that
the baryon asymmetry does not disappear and the baryon excess
partially
transforms into supersymmetric particles.
This is a reflection of the fact that  $B+L-R_2$  is anomalous,
but $B+L+R_2$ is anomaly-free.

So far we have assumed unbroken supersymmetry
and no explicit Higgsino mass terms $\mu H\bar H$ in the
superpotential. Gaugino Majorana masses
$M_{\tilde g},M_{\tilde W},M_{\tilde B}$ and soft trilinear scalar
couplings  explicitely
break $R_2$ and $R_3$, and the same is true for Higgsino
masses $\mu $. Let us take  all these soft terms
equal to a single symmetry-breaking parameter $M_{SS}\sim 10^2$
GeV. If  one    compares   the rate for these
symmetry-breaking effects,
$\Gamma _{SS} \simeq  {{M_{SS}^2}\over T}$
with the expansion rate of the universe $\Gamma _H\simeq 30\times
T^2/M_{Planck}$, one finds that the $R_{2,3}$ and
supersymmetry-breaking
effects are outside thermal equilibrium for
$T \geq  T_{SS} \simeq  {1\over {30^{1/3}}}\ M_{SS}^{2/3}
M_{Planck}^{1/3}  \simeq \ 10^7\ GeV.$
Thus above this temperature the arguments given in the previous
section apply.
Below that temperature, the explicit gaugino and Higgsino masses
force  the gaugino chemical potentials to vanish and the results
of previous analyses  \hatu, are recovered.

Let us  now   consider how the present analysis is modified in the
presence of extra non-renormalizable interactions violating
$B$ and/or $L$ symmetries \iq. We will only illustrate
 a particularly interesting case, that of the operator
%
${O_{\nu }\ =\ {1\over M}\ (L_LL_L\bar H\bar H )_F \ ,}$
which gives rise to Majorana neutrino masses upon
electroweak symmetry breaking of order
$m_{\nu }\simeq <\bar H>^2/M$. If $O_{\nu }$ is in thermal
equilibrium we will have the
extra chemical potential constraint
${L_L\ +\ \bar H \ =\ 0\ \ .}$
Above the $T_{SS}$ temperature one finds then that
$\tilde g\ =\ -{{99}/     {59}}\ Q_L
$
and there is only one independent chemical potential, e.g.  $Q_L$.
Then one has $(B+L)\propto Q_L\propto (B-L)$, and as long as there
was (or is created) a $B-L$ excess, the $B$ asymmetry is not
erased. Below $T_{SS}$,  $\tilde g=0$ and the baryon
asymmetry disappears unless the $O_{\nu }$ interaction
gets outside thermal equilibrium. Imposing that condition
one gets
$M\ \ge \ { 1\over {30^{2/3}}}\times M_{Planck}^{2/3}\times
M_{SS}^{1/3}\ \simeq \ 10^{12}\ GeV $
which in turn corresponds to a limit on (the heaviest) neutrino
mass
$m_{\nu }\ \le \ 10\ eV\ .$
This is to be compared with the much stronger limit \hatu,
  $m_{\nu }\le 10^{-3}$ eV, obtained  ignoring the existence
of the additional global currents.
Physically these four orders of magnitude are very important, as
mentioned in the introduction.

\medskip\bigskip\medskip

\centerline{\bf Gravitational Anomalies and $B-L$ Violation}
\medskip\bigskip
In the previous sections we were considering the fact that $B+L$ was
anomalous in the standard model but $B-L$  was exactly conserved.
Nevertheless it is straightforward to see that, in the
absence of right handed neutrinos, $B-L$ does have
anomalies when gravity is considered. The triangle diagram  couples
the $B-L$ current with two external gravitons instead of $SU(2)_W$
gauge fields.
It is then natural to inquire if there is a physical consequence to
this fact  in relation to the baryon asymmetry. We have seen
that (even in supersymmetric models) the nonperturbative electroweak
 effects
can erase any primordial $B+L$ asymmetry and we are left with  two
possibilities for baryogenesis. Either there is a primordial $B-L$
asymmetry or there is low energy baryogenesis due to  the same
electroweak
effects. The latter has been discussed by A. Cohen in this conference.
 The former could be achieved in different ways like decays of heavy
  (GUT)
particles, condensation of squarks and sleptons etc. Here we propose
 that
in the same spirit as low energy baryogenesis, the anomalous $B-L$
symmetry breaking could be the source of the baryon number asymmetry
of the universe. The scenario would be that gravitational anomalies
through nontrivial field configurations, induce a  $B-L$ asymmetry at
 temperatures of the order of $M_{Planck}$  or lower. The departure
 from thermal
equilibrium required for $B-L$ number generation would be given by
 the fact that
gravitational interactions become outside thermal equilibrium below
 $M_{Planck}$.
We leave unspecified a possible source of $CP$ violation.

  In the electroweak case, the mediators of $B+L$ breaking
 can be either Yang--Mills instantons (leading to a tunneling
 process which is exponentially suppressed) or sphalerons
 (providing
 the height of the energy barrier which can be passed at high
 temperatures).
 In the gravitational case, there are
 known field configurations resembling both. Gravitational instantons
 are
 solutions of euclidean Einstein's equations.
 The known, non--compact solutions are the
 asymptotically
  locally
 euclidean spaces. Their boundary has the topology of $S^3/G$ where
 $G$ is a
 discrete subgroup of $SU(2)$, therefore they have an $ADE$
 classification.
 The $A_k$ series corresponds to $G=\IZ _{k+1}$.
 The simplest of these spaces is the Eguchi--Hanson instanton
 ($k=1$)\eh. All of these manifolds are Kahler. They  have vanishing
 action, which is very different
 to the gauge case, in particular their effects are supposed to be
 proportional to $e^{-S}$ and there is no exponential suppresion
 if $S=0$. This fact seems to make these configurations more
 interesting than
 the corresponding gauge theory instantons, which have negligible
 effects for baryogenesis since $S\sim 1/g^2$. However all of these
 manifolds do not support spin--$1/2$ zero modes ($I_{1/2}=0$)
  when only gravitational
 backgrounds are considered. Nevertheless,  if there is also a
 $U(1)$ (or Yang--Mills) gauge field background \ \eh,\thd, there are
 spin--$1/2$
 zero modes but the action in this case is nonvanishing,
 \eqn\act{S={1\over{4e^2}}\int d^4x \sqrt{-g} F_{\mu\nu}^2=
 {4\pi^2\over e^2}N^2={\pi \over\alpha}N^2,}
 where $N\in \IZ$ is the instanton charge.
  Therefore these
 effects are also exponentially suppressed.  The natural candidate for
 the $U(1)$ is the standard model hypercharge. However,
  since $B-L$ has no mixed anomal
  y with hypercharge, no effect is
  to be expected in this case.   An alternative is to consider models
  with an extra $\t U(1)$ and such
  that this new interaction has a mixed anomaly with $B-L$.
  Examples of this type of $\t U(1)$'s are found in some supersymmetric
   models.
In this scheme, the gravitational piece does not contribute
directly to the asymmetry, but it  provides a background
to allow the $\t U(1)$ field develop a nontrivial topological
charge ($\int d^4x F\t F$), something which cannot happen in flat
space.
Also, as emphasized in \thd, the scale of these solutions can be
chosen large compared with $M_{Planck}$ allowing the possibility
to overcome the  problem of
dealing with gravity at the Planck scale. We may even speculate that
 in analogy to the electroweak case, the rate
for $B-L$ violating
transitions could go like $\Gamma\sim (\alpha T)^4$ which leads to
thermal equilibrium at energies $T \leq  \alpha^4 M_P\sim 10^{14}
Gev$. These low--energy effects would be interesting to
avoid a conflict with the inflationary universe scenario,
since if
  gravitational anomalies  create
a $B-L$ asymmetry at the Planck scale, inflation is expected
to dilute
 it
at lower energies and make it irrelevant.
 It would be
 very interesting if gravity could actually lead to such low--energy
 implications.

For this scenario to be realized, the analogous of the
`sphaleron' in the electroweak theory may be needed. It is
interesting
to notice that similar field configurations have been recently
found \bm\ in which a static (spherically symmetric) metric and
Yang--Mills backgrounds are unstable solutions of
Einstein--Yang--Mills equations with topological charge in between
two
stable vacua. Again the pure gravity part does not contribute to the
anomaly because of spherical symmetry ($\int d^4x  R\t R=0$) and it is the
gauge
 field the one that gives a nonvanishing contribution. These solutions,
 however,
 do not exist for the Einstein--Maxwell case. Also it is not clear if
 these configurations are physically identical to the standard model
 sphalerons. In particular, in gravity it is not well defined what
 the `height of the barrier' would actually be.

Gravitational anomalies effects have also been discussed in \gibbons\
in
the context of inhomogeneous cosmologies like Bianchi--IX. These
cosmologies give rise to the creation of neutrinos rather than
antineutrinos, the
asymmetry is related to the Atiyah--Patodi--Singer index for
manifolds with boundary. The number of neutrinos was found to be
$n_\nu\sim \exp{(12\beta)}/256$ where $\beta$ measures the initial
anisotropy of the space. This is different from our proposal since
in that case the lepton asymmetry is put, in some way, by hand by choosing
a sufficiently asymmetric initial cosmology. On the other hand, when the idea
in ref.\gibbons\ was put forward it was not known  that electroweak
effects could convert the initial lepton asymmetry into a baryon asymmetry.
We believe that  these issues should be reconsidered in the light of the
results
obtained in the last few years.
\medskip\bigskip

\centerline{\bf Acknowledgments}
\medskip\medskip

We thank L. Alvarez--Gaum\'e, E. Alvarez and A. Cohen for discussions
on the subject of the last section.
Research of FQ supported by the Swiss National
Science Foundation.

\listrefs
%

%
\bye